\documentclass[prl,aps,twocolumn,floatfix,showpacs]{revtex4}
\usepackage{graphicx,psfrag,amsmath,calc,amssymb,rotating}
\begin{document}

\title{Topological quantum phase transitions of ultracold fermions in optical lattices}

\author{R. W. Cherng}
\affiliation{Department of Physics, Harvard University,
Cambridge, MA 02138}

\author{C. A. R. S\'{a} de Melo}
\affiliation{School of Physics, Georgia Institute of Technology,
             Atlanta Georgia 30332}

\date{\today}

\begin{abstract}
We consider the possibility of topological quantum phase transitions of ultracold fermions 
in optical lattices, which can be studied as a function of interaction strength
or atomic filling factor (density). The phase transitions are connected to 
the topology of the elementary excitation spectrum, and occur only for non-zero angular momentum
pairing (p-wave, d-wave and f-wave), while they are absent for s-wave. 
We construct phase diagrams for the specific example of highly anisotropic optical lattices, 
where the proposed topological phase transitions are most pronounced. 
To characterize the existence of these topological transitions, 
we calculate several measurable quantities including momentum distribution, 
quasi-particle excitation spectrum, atomic compressibility, superfluid density,
and sound velocities.
\end{abstract}

\maketitle

In the last few years there has been a tremendous interest in atoms
loaded into optical lattices, where many known phases of standard condensed matter
physics can be simulated under a tighly controlled environment. 
Examples of such realizations are the observation of the superfluid-to-insulator~\cite{bloch-review-2007} 
transition in atomic Bose systems, metallic~\cite{esslinger-2005a}, molecular~\cite{esslinger-2006}
and superfluid~\cite{ketterle-2006} phases of atomic Fermi systems.
The production of quantum degeneracy of atoms and molecules in optical lattices has allowed the merger of 
standard condensed matter and atomic and molecular physics into a vibrant field of research called
condensed atomic and molecular physics.

In standard fermionic condensed matter systems it has been very difficult to study systematically
the effects of strong correlations as a function of particle density and the ratio of hopping to interaction
strength, as the ability to control these parameters is generally very limited. However, in atomic 
systems, the strength of atom-atom interactions, the atom hopping, and 
the atomic filling factor in optical lattices are easily controllable and allow 
the exploration of a much wider phase space. 
In particular, the ability to control interactions relies on the existence of Feshbach resonances, which have been used
to study fermions loaded into optical lattices both in the s-wave channel~\cite{ketterle-2006}
for $^6$Li, and the $p$-wave channel~\cite{esslinger-2005b} for $^{40}$K. 
More recently, $p$-wave Feshbach resonances in ${^6}$Li were used to produce 
p-wave molecules~\cite{fuchs-2008, inada-2008} in harmonically trapped systems, and it 
is likely that similar experiments may be attempted in optical lattices.

Currently, the only fully confirmed fermion superfluid with pairing higher than s-wave,
is liquid ${^3}$He, which is not found on a lattice. Very few lattice condensed matter 
candidates exist, such as the ruthenates~\cite{maeno-review-2000}, and 
organics~\cite{wei-review-2007},
but since the control of carrier density, hopping and interactions in these systems
is very limited, the exploration of the phase diagram of alleged $p$-wave
lattice superfluids has been hindered. Thus, the search of higher-angular momentum 
superfluidity in atomic systems is very important not only to help ellucidate the symmetry of the 
order parameter in analogous condensed matter systems, but also to explore new phases that 
are not accessible in standard condensed matter due to the lack of control of interactions,
density and hopping.

Since Feshbach resonances in optical lattices 
have already been observed in the $p$-wave channel~\cite{esslinger-2005b},
the interaction strength can be tuned continuously from weak (BCS) to strong (BEC) 
attraction limits.
Here, we take advantage of the tunability in optical lattices to 
study theoretically  
the occurance of unusual topological quantum phase transitions in three-dimensional 
but anisotropic optical lattices with superfluid order parameters in 
non-$s$-wave channels. 
We construct the phase diagram for anisotropic three-dimensional optical lattices
in the fermion density versus interaction strength plane and identify up to 
five different quantum phases for non-zero angular momentum states
depending on the momentum space topology of the quasiparticle excitation spectrum.
For some $p$-wave states, we find that the quasiparticle excitations are gapless in 
the BCS regime, and are fully gapped in the BEC regime. 
To characterize the change in topology of the quasiparticle excitations,
we show that the momentum distribution, atomic compressibility, superfluid density 
and sound velocity are non-analytic functions 
of the interaction strength exactly where the topological changes occur.

{\it Hamiltonian:} To describe the physics described above, we study ultracold fermions 
in anisotropic three-dimensional optical lattices described by the single band dispersion 
\begin{equation}
\label{eqn:dispersion}
\epsilon_{\bf k} = - 2t_x \cos(k_x a) - 2t_y \cos(k_y a) - 2t_z \cos(k_z a),
\end{equation}
Here, the hoppings $t_x > t_y > t_z $ are chosen to be different and $a$
is the optical lattice spacing.
We work with the Hamiltonian
$
H = H_{kin} + H_{int},
$
where the kinetic energy part is 
$
H_{kin} = \sum_{{\bf k},\alpha} \xi_{\bf k}
\psi_{{\bf k}, \alpha}^{\dagger} \psi_{{\bf k}, \alpha},
$
with $\xi_{\bf k} = \epsilon_{\bf k} - \mu$, where the
chemical potential $\mu$ may contain the standard Hartree shift. 
The interaction part of the Hamiltonian is
\begin{equation}
\label{eqn:hint}
H_{int} = 
\dfrac{1}{2} \sum_{{\bf k} {\bf k^{\prime}} {\bf q}} 
\sum_{\alpha \beta \gamma \delta}
V_{\alpha \beta \gamma \delta} ({\bf k}, {\bf k^{\prime}})
b_{\alpha \beta}^{\dagger} ({\bf k}, {\bf q})
b_{\gamma \delta} ({\bf k^{\prime}}, {\bf q})
\end{equation}
with
%
$
b_{\alpha \beta}^{\dagger} ({\bf k}, {\bf q}) = 
\psi_{-{\bf k} + {\bf q}/2, \alpha}^{\dagger}
\psi_{{\bf k} + {\bf q}/2, \beta}^{\dagger},
$
%
where the labels $\alpha$, $\beta$, $\gamma$ and $\delta$ 
are the pseudo-spin indices and 
the labels ${\bf k}$, ${\bf k}^{\prime}$ and ${\bf q}$ represent 
linear momenta. We use units where $\hbar = k_B = 1$, and allow
the pseudo-spins indices to take two values (pseudo-spin $S = 1/2$)
corresponding to two hyperfine states labeled as
$\vert\uparrow\rangle$ and $\vert\downarrow\rangle$.

In the case, where the hyperfine states (pseudo-spin indices) and the
center of mass coordinates are uncoupled the model interaction tensor 
can be chosen to be
\begin{equation}
V_{\alpha \beta \gamma \delta} ({\bf k}, {\bf k^{\prime}})
= - V_{\Gamma} 
\phi_{\Gamma} ({\bf k}) \phi^{*}_{\Gamma} ({\bf k^{\prime}})
\Gamma_{\alpha \beta \gamma \delta},
\end{equation}
where the tensor $\Gamma_{\alpha \beta \gamma \delta} = {\delta}_{\alpha {\bar \beta}}
{\delta}_{\gamma {\bar \delta}}^{\dagger}/2$, 
for the pseudo-singlet pairing case ($S_{1,2} = S_1 + S_2 = 0$); and where
$\Gamma_{\alpha \beta \gamma \delta} = {\bf v}_{\alpha \beta} \cdot
{\bf v}_{\gamma \delta}^{\dagger}/2$ with 
${\rm v}_{\alpha \beta} = (i\sigma \sigma_y)_{\alpha \beta}$,
for the pseudo-triplet pairing case ($S_{1,2} = S_1 + S_2 = 1)$. 
Here, $V_{\Gamma}$ has dimensions of energy and represents 
a given symmetry of the order parameter with basis function
$\phi_{\Gamma} ({\bf k})$ and $\phi_{\Gamma}^{*} ({\bf k^\prime})$ 
representative of the orthorhombic group ($D_{2h}$).

{\it Self-Consistent Equations:} 
At the saddle point, the pairing field ${\cal D}_{\lambda} ({\bf k_1} + {\bf k_2}, \tau)$ is taken to be 
$\tau$ independent, and to have center of mass momentum ${\bf k_1} + {\bf k_2} = 0$, 
becoming 
$
{\cal D}_{0} ({\bf k_1} + {\bf k_2}, \tau) = 
\Delta_{\Gamma} \delta_{ {\bf k_1} + {\bf k_2}, 0} + 
\delta {\cal D}_{0} ({\bf k_1} + {\bf k_2}, \tau) 
$
for the singlet case, and
$
{\cal D}_{i} ({\bf k_1} + {\bf k_2}, \tau) = 
\eta_i \Delta_{\Gamma} \delta_{ {\bf k_1} + {\bf k_2}, 0} + 
\delta {\cal D}_{i} ({\bf k_1} + {\bf k_2}, \tau) 
$
for the triplet case. 
The corresponding order parameter equation is
\begin{equation}
\label{eqn:order-parameter}
1 = \sum_{\bf k} \vert V_{\Gamma} \vert |\phi_{\Gamma} ({\bf k})|^2 
\tanh (\beta E_{\bf k}/2) / 2 E_{\bf k},
\end{equation}
The number equation is obtained from 
$N = - \partial \Omega/\partial \mu$, 
where $\beta \Omega  = - \ln Z$ is
the thermodynamic potential and $Z = {\rm Tr} (\exp{-\beta H})$
is the partition function, leading to 
\begin{equation}
\label{eqn:number}
N = N_0 + N_{\rm fluct},
\end{equation}
where $N_0 = \sum_{\bf k} n_{\bf k}$, 
and $n_{\bf k} = 
\left[
1 - \xi_{\bf k} \tanh (\beta E_{\bf k}/2) / E_{\bf k} 
\right]
$
is the momentum distribution.
The additional term $N_{\rm fluct} = - \partial \Omega_{\rm fluct}/\partial \mu$, 
where $\Omega_{\rm fluct}$ are Gaussian fluctuations to saddle point $\Omega_0$.
These two equations must be solved self-consistently in order to provide 
the order parameter amplitude $\Delta_\Gamma$, the chemical potential $\mu$,
and the quasiparticle excitation energy
$$
E_{\bf k} = 
\sqrt{ \xi_{\bf k}^2 + |\Delta_{\Gamma}|^2 |\phi_{\Gamma} ({\bf k})|^2}.
$$

{\it Order Parameter Symmetries:} For singlet pairing the saddle point field 
is ${\cal D}_0^{(0)} = \Delta_\Gamma$, while
for the triplet case it is ${\cal D}_i^{(0)} = \eta_i \Delta_\Gamma$, which is related 
to the ${\bf d}$-vector order parameter by 
$
{d_i} ({\bf k}) = 
\sum_{\bf k} {\cal D}_i^{(0)} \phi_{\Gamma} ({\bf k}).
$
For orthorhombic lattices without breaking time-reversal or parity 
and no coupling between pseudospins and center of mass coordinates, 
the only order parameters for superfluidity allowed by symmetry are:
(a) $\Delta ({\bf k}) = \Delta_\Gamma \phi_{\Gamma} ({\bf k}) $ for
singlet states and 
(b) ${\bf d} ({\bf k}) = {\hat \eta} \Delta_{\Gamma} \phi_{\Gamma} ({\bf k}) $ 
for triplet states. 
This implies that there are only 8 symmetries allowed for the order parameter
which are consistent with the orthorhombic $D_{2h}$ group. 
There are four options for the singlet case:
(a) $s$-wave with $\Delta ({\bf k}) = \Delta_s$;
(b) $d_{xy}$-wave with $\Delta ({\bf k}) = \Delta_{d_xy} XY$;
(c) $d_{xz}$-wave with $\Delta ({\bf k}) = \Delta_{d_xz} XZ$; and
(b) $d_{yz}$-wave with $\Delta ({\bf k}) = \Delta_{d_yz} YZ$.
There are also four options for the triplet case:
the ${\bf d}$-vector in momentum space for unitary triplet
states in the weak spin-orbit coupling limit
is characterized by one of the four states:
(a) $p_x$-wave with ${\bf d} ({\bf k}) = {\hat \eta} \Delta_{p_x} X$;
(b) $p_y$-wave with ${\bf d} ({\bf k}) = {\hat \eta} \Delta_{p_y} Y$;
(c) $p_z$-wave with ${\bf d} ({\bf k}) = {\hat \eta} \Delta_{p_z} Z$; and
(d) $f_{xyz}$-wave with ${\bf d} ({\bf k}) = {\hat \eta} \Delta_{f_{xyz}} X Y Z$. 
Since, the Fermi surface can touch the Brillouin zone boundaries the functions 
$X$, $Y$, and $Z$ need to be periodic and can 
be chosen to be $X = \sin{(k_{x} a)}$, 
$Y = \sin{(k_{y} a)}$, 
and $Z = \sin{(k_{z} a)}$.
The unit vector $\hat \eta$ defines the direction 
of ${\bf d} ({\bf k})$.
From here on we scale all energies by $t_x$,and choose  
$t_y/t_x = 0.2$ and $t_z/t_x = 0.008$ such that $t_x \gg t_y \gg t_z$ 
where the largest number of non-trivial phases occur.

\begin{figure}
\begin{center}
\(
  \begin{array}{c}
   \multicolumn{1}{l}{\hspace{-0.2cm}\mbox{\bf (a)}} \\
\psfrag{N}{$\widetilde N$}
\psfrag{V}{$\vert V_{p_y}/2 t_x \vert $}
\psfrag{m1}{$\mu_1^*$}
\psfrag{m2}{$\mu_4^*$}
\includegraphics[width=6.2cm]{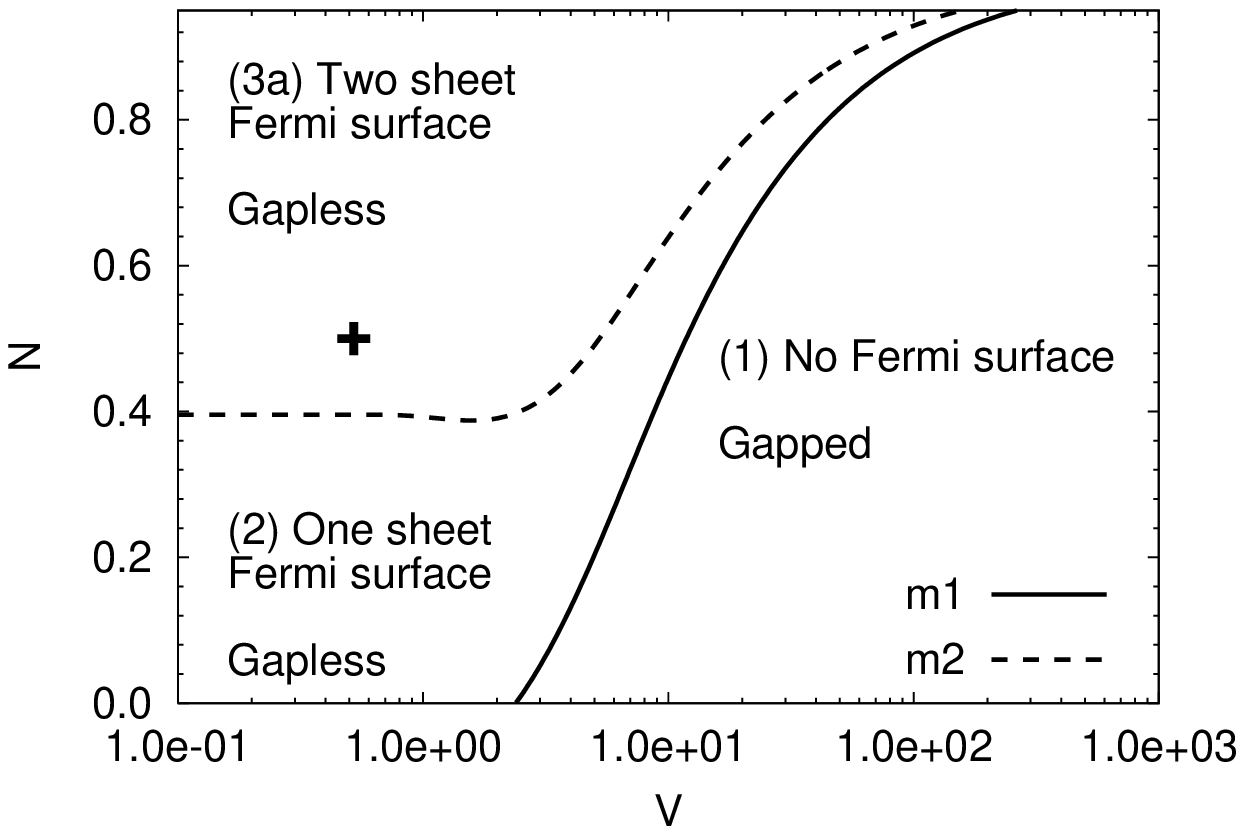} \\

   \multicolumn{1}{l}{\hspace{-0.2cm}\mbox{\bf (b)}} \\
\psfrag{z}{\begin{sideways}
{\begin{sideways} \Huge $T/\vert 2 t_x \vert$ \end{sideways} } 
\end{sideways}}
\psfrag{x}{\Huge  $\vert V_{p_y}/2 t_x \vert$}
\centerline{\scalebox{0.50}{\includegraphics{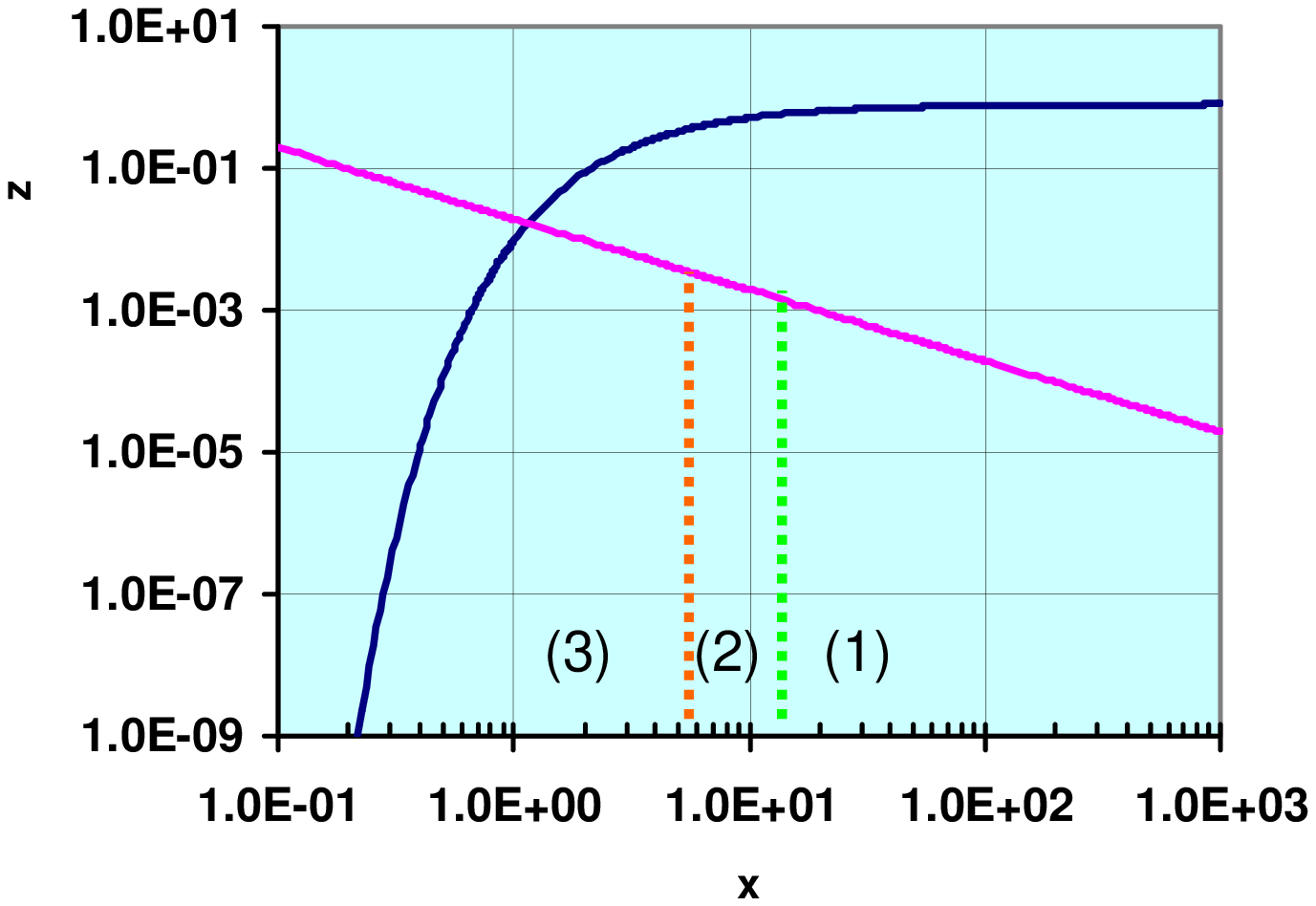} } } 
\end{array}\)
\end{center}
\caption{a) The T = 0 phase diagram of filling 
factor ${\widetilde N}$ versus interaction $V_{p_y}$.
The cross indicates the location of a suspected 
$p$-wave superfluid in highly anisotropic lattices 
of standard condensed matter, see Ref.~\cite{wei-review-2007}.
b) the temperature versus interaction phase diagram 
at ${\widetilde N} = 0.5$ (quarter filling) for the $p_y$ symmetry. 
The solid lines indicate asymptotic forms of the critical
temperatures for weak and strong coupling, and the dotted lines indicate 
the phase boundaries between topological phases of 
the type (1), (2) and (3). The hopping ratios used are 
$t_y/t_x = 0.2$, and $t_z/t_y = 0.008$.
}
\label{fig:phase-diagram}
\end{figure}

{\it Topological Transitions:} We discuss three distinct phases based on 
the normal state and quasiparticle Fermi surface (FS) topologies and 
the nodal structure of the order parameter as a function of $\mu$.
We consider the ``normal state'' FS defined in the first Brillouin zone by  
$\xi_k=0$, keeping in mind the periodicity in $k$-space, and define the 
special values
$\mu^*_1\equiv - 2t_x - 2t_y - 2t_z$; 
$\mu^*_2\equiv - 2t_x - 2t_y  + 2t_z$; 
$\mu^*_3\equiv - 2t_x + 2t_y - 2t_z$; 
$\mu^*_4\equiv - 2t_x + 2t_y + 2t_z$ 
for filling factors $0 \le \widetilde N \le 1$.
The order parameter always has no nodes for the $s$ symmetry, while the nodes for the $p_i$ symmetry
can only occur on the planes $k_i=0,\pm\pi$ where $i$ is $x$, $y$, or $z$.  
For the superconducting state, the intersection of the FS and order parameter nodes constitute
the loci of gapless excitations, where $E({\bf k}) = 0$. 
Since $E({\bf k})$ is always gapped for the $s$-wave symmetry, there are no quantum phase transitions
present. However, topological quantum phase transitions occur for $p_i$ symmetries.

For all triplet $p_i$-wave symmetries, $E({\bf k})$ is fully gapped only for 
$\mu < \mu^*_1$ since there is no Fermi surface. 
For $\mu^*_1 < \mu < \mu^*_4$, the order parameter nodes intersect the 
Fermi surface and hence quasiparticle excitations are gapless.
For $\mu^*_4 < \mu$ the Fermi surface splits into two sheets that separate along $k_x$ so that
the $p_x$ nodes no longer intersect the Fermi surface opening a gap in the excitation spectrum. 
However, the $p_y$, and $p_z$ nodes still intersect the Fermi surface so excitations
remain gapless. 

Representative phase diagrams for the $p_y$ symmetry are shown in 
Fig.\ref{fig:phase-diagram}. 
There are three distinct superfluid phases 
characterized by Fermi surface connectivity and quasiparticle excitation spectrum:
(1) no Fermi surface and fully gapped for all $p_i$ symmetries ($\mu < \mu^*_1$),
(2) one sheet Fermi surface and gapless for all $p_i$  symmetries ($\mu^*_1 < \mu < \mu^*_4$),
(3a) two sheet Fermi surface and gapless for $p_y$, $p_z$  ($\mu^*_4 < \mu$)
(3b) two sheet Fermi surface and fully gapped for $p_x$ ($\mu^*_4 < \mu$).
In addition, phase (2) splits into three regions using the finer classification of 
Fermi surface topological genus:
(2i) genus zero ($\mu^*_1 < \mu < \mu^*_2$), (2ii) genus one ($\mu^*_2 < \mu < \mu^*_3$),
(2iii) genus two ($\mu^*_2 < \mu < \mu^*_3$).
There are several qualitative features of interest in the phase diagrams.  For a fixed low density
and all triplet symmetries, the chemical potential does not go below the bottom of the band
until a critical coupling is reached indicating the formation of a bound state requires a
finite interaction strength, which occurs at
$\vert V_{p_x}/2t_x \vert  = 3.5052$, $\vert V_{p_y}/2 t_x \vert = 2.3952$, 
and $\vert V_{p_z}/2 t_x \vert = 1.8150$, 
for the $p_x$, $p_y$, and $p_z$ symmetries respectively.

\begin{figure}
\begin{center}
\(
  \begin{array}{c}
   \multicolumn{1}{l}{\hspace{-0.2cm}\mbox{\bf (a)}} \\
     \includegraphics[width=6.0cm]{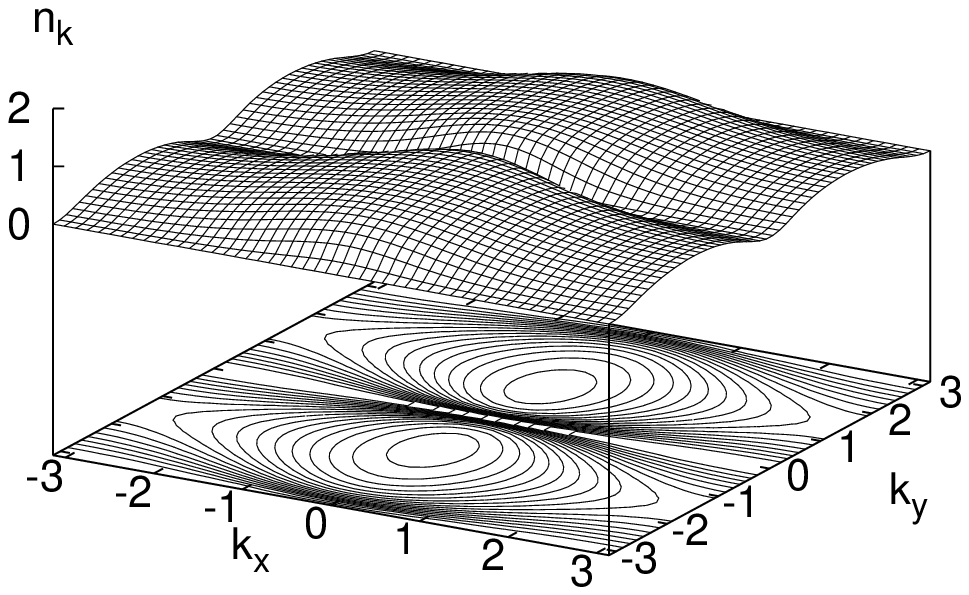} \\
   \multicolumn{1}{l}{\hspace{-0.2cm}\mbox{\bf (b)}} \\
     \includegraphics[width=6.0cm]{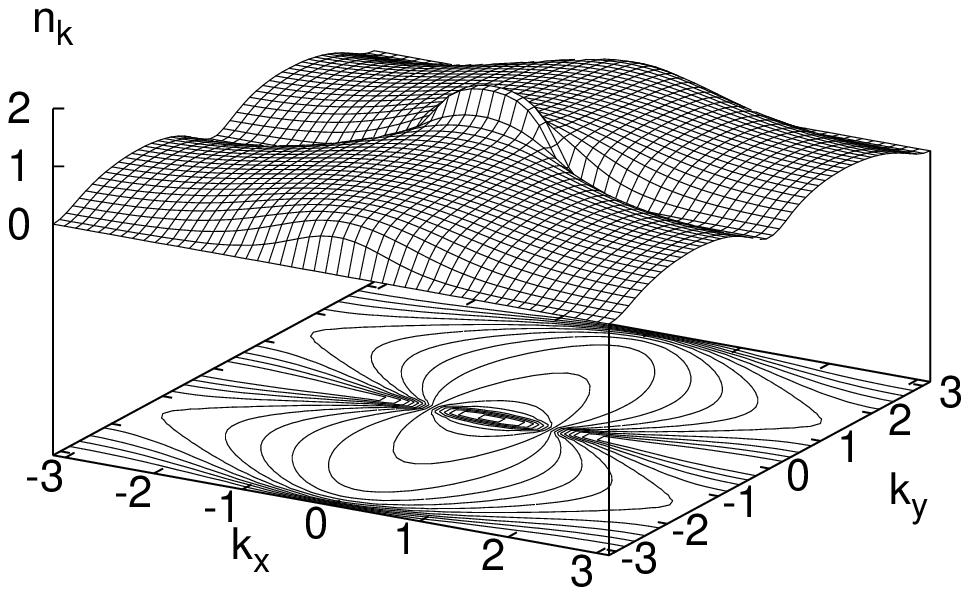}
\end{array}\)
\end{center}
\caption{Momentum distribution in the plane $k_z=0$ for the $p_y $ symmetry
for ${\widetilde N} = 0.5$ and in the vicinity of $\vert V_{p_y}/2 t_x \vert \approx 11$  
when (a) $\mu < \mu^*_1$ and (b) $\mu > \mu^*_1$. 
Note the jump in the momentum distribution at the origin from a) to b),
signaling a topological transition at $\mu = \mu_1^*$.} 
\label{fig:nk-plot}
\end{figure}

{\it Momentum Distribution:} While changes for $s$-wave are smooth, dramatic 
rearrangements of the momentum distribution $n ({\bf k})$ accompany 
the topological transitions for $p$-wave $d$-wave and $f$-wave. 
For brevity, we analyze only the $p_i$ symmetries 
where $i=x,y,z$. Near the points where the Fermi surface changes 
topology (connectivity or genus) 
$
n ({\bf k} ) \approx 1 + {\rm sgn}(\delta\mu) - {\rm sgn}(\delta\mu) 
\Delta_{p_j}^2 \vert \delta {\bf k}_j \vert^2 / ( 2 \vert \delta\mu \vert ^2 )
$
where $\delta \mu = \mu - \mu^*_j$ and 
$\delta {\bf k} = {\bf k}- {\bf k}^*_j$ where $ j = 1,2,3,4 $.  The momentum ${\bf k}_j^*$ 
is the point (up to lattice periodicity) where the Fermi surface changes topology: 
${\bf k}_1^* = (0,0,0)$, ${\bf k}_2^* = (0,0,\pi)$, ${\bf k}_3^* = (0,\pi,0)$, 
${\bf k}_4^* = (0,\pi,\pi)$. This expansion clearly breaks down for $\delta \mu = 0$, 
where we find
$n ({\bf k}) \approx 
1- \vert t_j \vert  \vert \delta {\bf k}_i \vert / \Delta_0 
$
for $\vert \delta {\bf k}_j \vert \ne 0 $, 
and 
$n ({\bf k}) \approx 0$ for 
$\vert \delta {\bf k}_j \vert = 0 $. 
When $\mu$ crosses the boundaries $\mu^*_j$, there are clear discontinuities in 
$n({\bf k})$. Plots of $n({\bf k})$ for the $p_y$ symmetry are shown in
Fig.~\ref{fig:nk-plot} for ${\widetilde N} = 0.5$ and in the vicinity of $\mu \approx \mu^*_1$ 
$(\vert V_{p_y}/2 t_x \vert \approx 11 )$. 

{\it Compressibility and Superfluid density:} The atomic compressibility and superfluid
density provide additional signatures of topological quantum phase transitions in optical lattices.
These quantities can be directly extracted from an analysis of quantum $(T = 0)$ phase fluctuations.
The effective action 
$S_{\rm eff} = S_{\rm eff}^{(0)} + \delta S_{\rm eff}$ has a saddle point 
$S_{\rm eff}^{(0)} = \beta \Omega_0$ and fluctuation
$\delta S_{\rm eff} = \beta \Omega_{\rm fluct}$ contributions. 
Integrating the amplitude fluctuations in $\delta S_{\rm eff}$ leads to the phase only action
\begin{equation}
\label{eqn:phase-action}
\delta S_{\rm phase} = \frac {1}{8} \sum_{q, \omega_n} \left[ A (\omega_n)^2 + 
\rho_{ij} q_i q_j \right] \theta (q) \theta (-q),
\end{equation}
with $\theta (q)$ being the phase field of the order parameter and  
$A = N^2 \kappa/V$, where 
$
\kappa = - \frac{1}{V} \left( \frac {\partial V} {\partial P} \right)_{T,N}
= \frac{1}{N^2} \left( \frac {\partial N} {\partial \mu} \right)_{T,V}
$
is the atomic compressibility
and $\rho_{ij} = V^{-1} \sum_{\bf k} \left[ n_{\bf k} \partial_i \partial_j \xi_{\bf k} \right]$
is the superfluid density tensor.
Notice that $\Delta S$ does not only give $\kappa$ and $\rho_{ij}$ but also the phase-only
collective mode (sound) velocity $\omega ({\bf k}) = \sqrt{ c_x^2 q_x^2 + c_y^2 q_y^2 + c_z^2 q_z^2 }$,
where $c_x^2 = \rho_{xx}/A$, $c_y^2 = \rho_{yy}/A$ and $c_z^2 = \rho_{zz}/A$, when the 
analytic continuation $i\omega_n \to \omega + i\delta$ is performed.

\begin{figure}
\begin{center}
\(
  \begin{array}{c}
   \multicolumn{1}{l}{\hspace{-0.2cm}\mbox{\bf (a)}} \\
\psfrag{k}{$ 2 \kappa \vert t_x \vert$}
\psfrag{V}{$\vert V_{p_y}/ 2 t_x \vert$}
     \includegraphics[width=6.0cm]{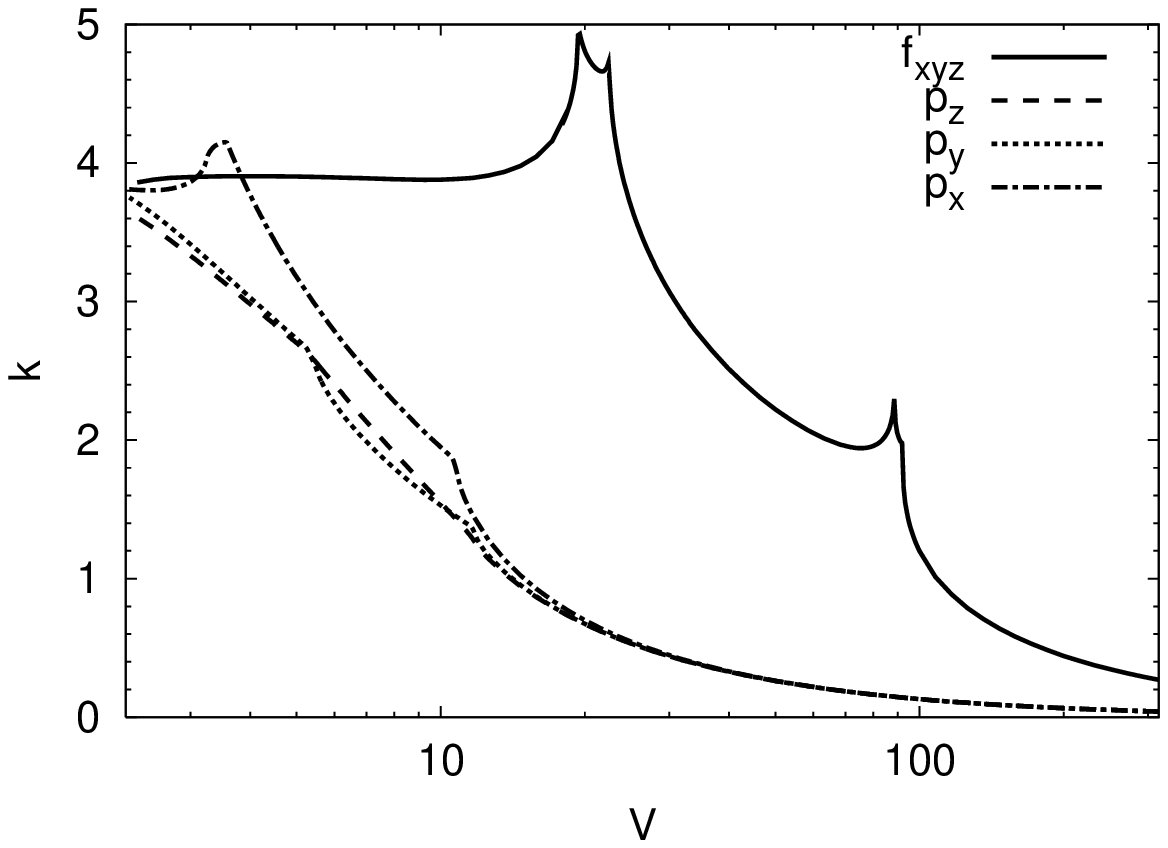} \\
   \multicolumn{1}{l}{\hspace{-0.2cm}\mbox{\bf (b)}} \\
\psfrag{cyy}{$c_{y}/ \vert t_x a \vert$}
\psfrag{V}{$ \vert V_{p_y}/2 t_x \vert$}
     \includegraphics[width=6.0cm]{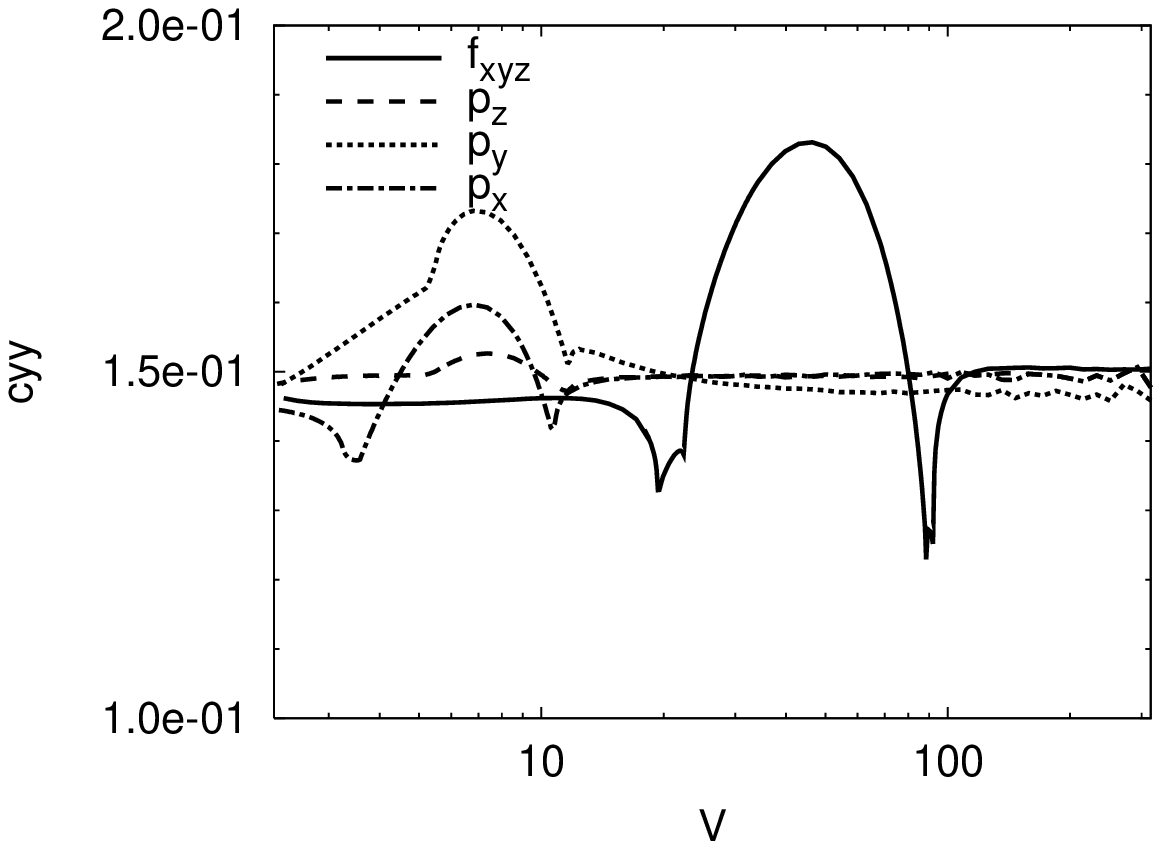}
\end{array}\)
\end{center}
\caption{ Plots of (a) atomic compressibility $\kappa$ and (b) sound velocity
$c_y$ versus interaction $V_{p_y}$ for the $p_y$-symmetry
at zero temperature. The parameters are the same as in Fig.~\ref{fig:phase-diagram} }
\label{fig:kappa-sound}
\end{figure}

Measurement of two quantities out of three 
(sound velocity, superfluid density and compressibility) can yield the
third one. In particular, the techniques used to measure accurately the sound 
velocity of superfluid fermions in harmonic traps~\cite{grimm-2004,thomas-2005}
and to measure compressibility~\cite{bloch-2008} can be used in the lattice case.
In Fig.~\ref{fig:kappa-sound}, we show (a) the atomic compressibility $\kappa$,  
and (b) the sound velocity $c_y = \sqrt{(\rho_{yy} V)/(N^2 \kappa)}$ 
along the $y$-direction. Notice the clear non-analytic behavior 
in these two quantities when the phase boundaries between
different topological phases are crossed. 

The non-analytic behavior of the momentum distribution, sound velocity and compressibility,
as function of interaction strength or density (not shown) characterizes 
the topological quantum phase transition described. 
Since topological transitions occur without a change in symmetry, 
they can not be classified under Landau's scheme, however discontinuities 
occur in the third derivative of the thermodynamic potential (first derivative
of compressibility), and the transition can be identified as a third-order transition
under Ehrenfest's classification. This type of topological quantum phase transition belongs 
to a class of transitions first envisioned by Lifshitz~\cite{lifshitz-1960}
for non-interacting Fermi systems, where the Fermi surface changes its topology of 
under the influence of external pressure. Lisfhitz's original idea was extended to continuum 
theories of interacting fermions by Volovik~\cite{volovik-1992} in the context of $^3$He, 
where topological invariants were constructed for topological changes in the quasiparticle 
excitation spectrum, but thermodynamic signatures were not analyzed. 
Here, we generalized their pioneering work to the case  
interacting fermions in a lattice, and described 
thermodynamic and transport signatures of topological 
quantum phase transitions in the superfluid phase.

{\it Summary:} We studied topological quantum phase transitions of ultracold fermions in
optical lattices. For brevity, we focused on quasi-one-dimensional optical lattices, 
where the nature of the topological transitions in the superfluid state is most dramatic. 
We classified the quantum superfluid phases in accordance to their Fermi surface topologies 
and quasiparticle excitation spectrum. 
We showed that for $s$-wave superfluids there
is no phase transition, but for $p$-wave ($d$-wave or $f$-wave) superfluids, quantum phase 
transitions of topological nature can occur, in particular from a phase with gapless quasiparticle
excitations (BCS regime) to a phase of with gapful quasiparticle excitations (BEC regime).
Finally, we showed that non-analytic behavior of the momentum
distribution, compressibility, sound velocities and superfluid density tensor 
as a function of interaction strength characterize well the topological phase transitions
for $p_x$, $p_y$ and $p_z$, and $f_{xyz}$ pairing symmetries. 
We would like to thank NSF for support(Grant No. DMR-0709584)

\end{document}